\documentstyle[amssymb,12pt,aps,prl,graphicx,epsfig]{revtex}

\begin{document}
\title{On a possible dynamical scenario leading to a generalised Gamma
distribution}
\author{Silvio M. Duarte Queir\'{o}s\thanks{%
Electronic address: {\tt sdqueiro@cbpf.br}}}
\address{Centro Brasileiro de Pesquisas F\'{i}sicas, 150, 22290-180 Rio de\\
Janeiro-RJ, Brasil}
\date{\today}
\maketitle

\begin{abstract}
In this report I present a possible scenario which can lead to the
emergence of a generalised Gamma distribution first presented by C Tsallis 
\textit{et al.} as the distribution of traded volumes of stocks in financial
markets. This propose is related with superstatics and the notion of moving
average commonly used in econometrics.
\end{abstract}

\section{The $\Gamma $-distribution}

\begin{figure}[tbp]
\begin{center}
\includegraphics[width=.75\columnwidth,angle=0]{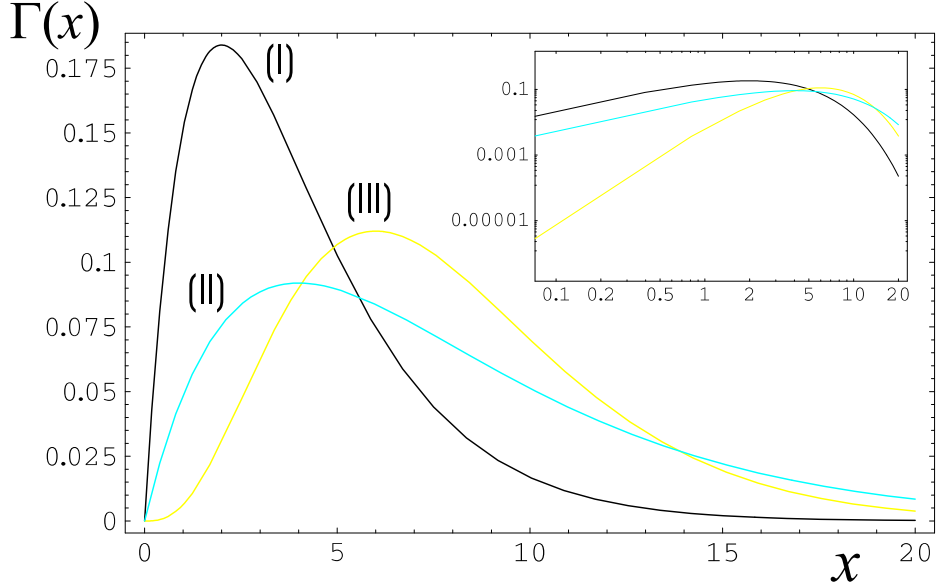}
\end{center}
\caption{Representation of $\Gamma \left(
x\right) $ $vs$ $x$ for some values of $\protect\alpha $ and $\protect\theta 
$. (I) - $\protect\alpha =2$ and $\protect\theta =2$;(II) - $\protect\alpha %
=4$ and $\protect\theta =2$; (III) - $\protect\alpha =2$ and $\protect\theta %
=4$. The inset presents the same curves, but in a $\log -\log $ scale.}
\label{fig-1}
\end{figure}

The $\Gamma $-distribution is a general distribution that is verified in
processes where the waiting times between variables that follow a Poisson
distribution are significant. It involves two free parameters, usually
labeled by $\alpha $ and $\theta $ and defined as \cite{jambu},
\begin{equation}
p_{\alpha ,\theta }\left( x\right) =\frac{x^{\alpha -1}\exp \left( -\frac{x}{%
\theta }\right) }{\Gamma \left[ \alpha \right] \,\theta ^{\alpha }}.
\label{def-gamma}
\end{equation}
A special case of $\Gamma $-distribution is to consider $\alpha =r/2$ and 
$\theta =2$. In this case the distribution is called $\chi ^{2}$-distribution
and represents the probability of get a value $\zeta $ of a variable that is
obtained by the summation of $r$ independent squared variables $\xi _{i}$
associated with the Gaussian distribution with null mean and unitary
variance \cite{jambu},
\begin{equation}
\zeta =\sum\limits_{i=1}^{r}\xi _{i}^{2}.
\label{def-gamma1}
\end{equation}
The same form presented in Eq. (\ref{def-gamma}) can be obtained as the
stationary solution of the following differential stochastic equation
\begin{equation}
dx_{t}=-\gamma \left( x_{t}-\theta \right) \,dt+k\sqrt{x}dW_{t}.
\label{langevin-gamma}
\end{equation}
Considering the It\^{o} convention for stochastic differentials I am able to
write the Fokker-Planck equation \cite{risken},
\begin{equation}
\frac{\partial p\left( x,t\right) }{\partial t}=\frac{\partial }{\partial x}%
\left[ \gamma \left( x_{t}-\theta \right) \,p\left( x,t\right) \right] +%
\frac{1}{2}\frac{\partial ^{2}}{\partial x^{2}}\left[ k^{2}\,x\,p\left(
x,t\right) \right] ,  \label{fokker-planck-gamma}
\end{equation}
whose stationary solution is
\begin{equation}
p_{\alpha ,\theta }\left( x\right) =\frac{\alpha ^{\alpha }}{\Gamma \left[
\alpha \right] }\frac{x^{\alpha -1}}{\theta ^{\alpha }}\exp \left( -\frac{%
\alpha }{\theta }x\right)   \label{gamma-yakovenko}
\end{equation}
with $\alpha =2\gamma \theta k^{-2}$. Performing a simple variable change 
$x\rightarrow \frac{x}{\alpha }$, it is possible to transform Eq. 
(\ref{gamma-yakovenko}) into Eq. (\ref{def-gamma}) and the It\^{o}-Langevin
equation (\ref{langevin-gamma})
\begin{equation}
dx=-\gamma \left( x-\alpha \,\theta \right) \,dt+\sqrt{2\,\gamma \,\theta }\sqrt{x}\,dW_{t}.
\label{langevin-normalised}
\end{equation}
For a question of simplicity let me represent $\theta $ as $\beta ^{-1}$. So
Eq. (\ref{def-gamma}) will be written as,
\begin{equation}
p_{\alpha ,\beta }\left( x\right) =\frac{\beta ^{\alpha }}{\Gamma \left[
\alpha \right] }x^{\alpha -1}\exp \left( -\beta \,x\right) .
\label{gamma-reescrita}
\end{equation}
In  figure \ref{fig-1} are depicted some examples of $\Gamma \left( x\right) $
distributions.

\section{Introducing the generalised $\Gamma $-distribution}

Let one now suppose that parameter $\theta $ in Eq. (\ref%
{langevin-gamma}) is in fact a stochastic variable on time scale larger than
the characteristic time scale $\gamma ^{-1}$. This means that $p_{\alpha
,\beta }\left( x\right) $ is, for this case, a conditional probability
density function $p_{\alpha }\left( x\,|\,\beta \right) $. If the random
process for $\beta $ is associatd with a SPDF, $\Pi \left( \beta \right) $,
then the SPDF for $x$ variable, $P\left( x\right) $, is simply given by
\begin{equation}
P\left( x\right) =\int p_{\alpha }\left( x\,|\,\beta \right) \,\Pi \left(
\beta \right) \,d\beta .  \label{prob-composta}
\end{equation}

Among the various distributions for non-negative variables let one
consider that $\beta $ is associated, itself, with a $\Gamma $-distribution,
\begin{equation}
P\left( x\right) =\int p_{\alpha }\left( x\,|\,\beta \right) \,\Pi \left(
\beta \right) \,d\beta .
\end{equation}
which can be associated to a microscopic equation similar to Eq. (\ref%
{langevin-normalised}).

Calculating the integral presented in equation (\ref{prob-composta}) one
gets,
\begin{equation}
P\left( x\right) =\frac{\Gamma \left[ \alpha +\lambda \right] \,\omega
^{\alpha }}{\Gamma \left[ \alpha \right] \,\Gamma \left[ \lambda \right] }%
x^{\alpha -1}\,\left( 1+\omega \,x\right) ^{-\left( \alpha +\lambda \right)
}.  \label{q-gamma-1}
\end{equation}

\begin{figure}[tbp]
\begin{center}
\includegraphics[width=.75\columnwidth,angle=0]{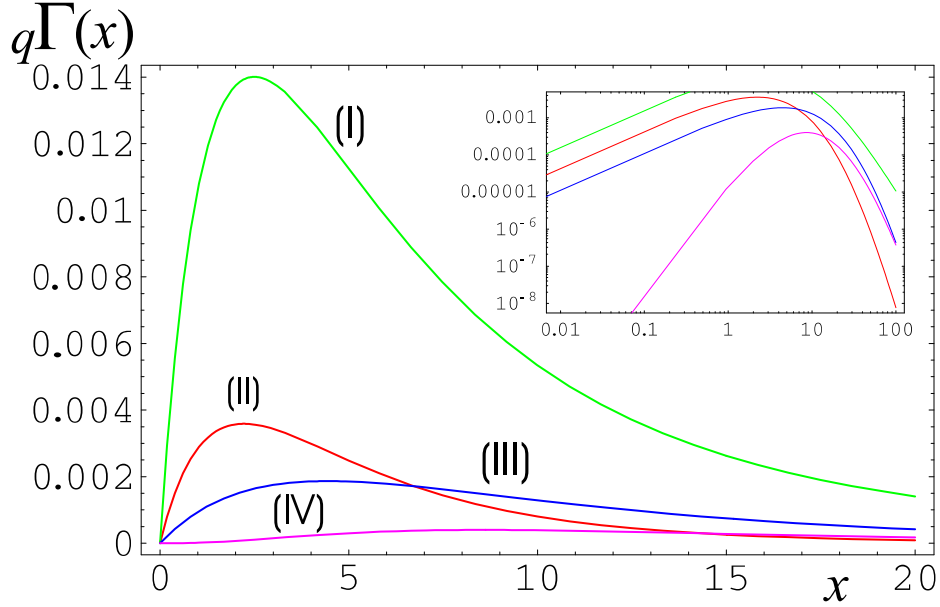}
\end{center}
\caption{Representation of $\Gamma \left(
x\right) $ $vs$ $x$ for some values of $\protect\alpha $ and $\protect\theta 
$. (I) - $q=1.2$, $\protect\alpha =2$ and $\protect\theta =2$; (II) - $q=1.1$%
, $\protect\alpha =2$ and $\protect\theta =2$;(III) - $q=1.1$,  $\protect%
\alpha =2$ and $\protect\theta =4$; (IV) - $q=1.1$, $\protect\alpha =4$ and $%
\protect\theta =2$. The inset presents the same curves, but in a $\log -\log 
$ scale.}. The same form is presented in Figs. 7 and 8 of Ref. \cite{osorio}.
\label{fig-2}
\end{figure}

Defining $\bar{\theta}=\frac{1}{\omega \,\left( \alpha +\lambda
\right) }$ and $q=1+\frac{1}{\alpha +\lambda }$, Eq. (\ref{q-gamma-1}) can
be rewritten as,
\begin{equation}
P\left( x\right) =\frac{\Gamma \left[ \frac{1}{q-1}\right] \,\left( \frac{q-1%
}{\bar{\theta}}\right) ^{\alpha }}{\Gamma \left[ \alpha \right] \,\Gamma %
\left[ \frac{1}{q-1}-\alpha \right] }x^{\alpha -1}\left\{ 1-\left(
1-q\right) \,\frac{x}{\bar{\theta}}\right\} ^{\frac{1}{1-q}},
\end{equation}
\begin{equation}
P\left( x\right) \equiv \frac{\Gamma \left[ \frac{1}{q-1}\right] \,\left( 
\frac{q-1}{\bar{\theta}}\right) ^{\alpha }}{\Gamma \left[ \alpha \right]
\,\Gamma \left[ \frac{1}{q-1}-\alpha \right] }x^{\alpha -1}\exp _{q}\left[ -%
\frac{x}{\bar{\theta}}\right] ,
\label{q-gamma-final}
\end{equation}
which I will call the $_{q}\Gamma $-\textit{distribution}. This kind of
distribution was already verified, at least, for the distribution of traded
volumes of stocks in financial markets \cite{osorio}. For the limit $q\rightarrow 1$, the
usual $\Gamma $-distribution is recovered, which corresponds to $\Pi \left(
\beta \right) =\delta \left( \beta -\frac{1}{\theta }\right) $. Some
examples of $_{q}\Gamma $-distribution are presented in Fig. \ref{fig-2},

This problem of fluctuations in some intensive parameter of the dynamical
equation(s) that describe(s) the evolution of a system \cite{wilk} was recently studied
by C. Beck in the context of Langevin equation with fluctuating temperature \cite{beck}
and extended together with Eddie G.D. Cohen \cite{beck-cohen} who defined it as \textit{%
superstatistics} (a statistic of statistics). This superstatistics presents
a close relation to the non-extensive statistical mechanics framework based
on the entropic form \cite{ct,gm-ct},
\begin{equation}
S_{q}=\frac{1-\int \left[ p\left( x\right) \right] ^{q}\,dx}{q-1}.
\label{s-q}
\end{equation}
For the problem of the distribution of traded volume of stocks in financial
markets the presence of fluctuations in $\theta $, or the mean value of the
scaled variable $\alpha \,x$, it is similar to the problem of the moving
average in the analysis of the volatility useful in the reprodution of some
empirical facts like the autocorrelation function and the so-called leverage
efect, see e.g. Ref. \cite{bouchaud}.

\end{document}